\documentclass[aps,prl,twocolumn,groupedaddress]{revtex4}
\usepackage{graphicx}

\begin{document}

%\preprint{}

\title{Nuclear Spin Switch in Semiconductor Quantum Dots}


\author{A.I. Tartakovskii$^{1}$, T. Wright$^{1}$, A. Russell$^{2}$, V.I.
Fal'ko$^{2}$, A. B. Van'kov$^{1}$, J. Skiba-Szymanska$^{1}$, I. Drouzas$^{1}$,
R.S. Kolodka$^{1}$, M.S. Skolnick$^{1}$, P.W. Fry$^{3}$, A. Tahraoui$^{3}$,
H.-Y. Liu$^{3}$, M. Hopkinson$^{3}$}

\affiliation{$^{1}$ Department of Physics and Astronomy, University of Sheffield, S3 7RH,UK\\
$^{2}$ Department of Physics, University of Lancaster, Lancaster LA1 4YB, UK\\
$^{3}$ Department of Electronic and Electrical Engineering, University of Sheffield, Sheffield S1 3JD, UK}

\date{\today}

\begin{abstract}
We show that by illuminating an InGaAs/GaAs self-assembled quantum dot with
circularly polarized light, the nuclei of atoms constituting the dot can be
driven into a bistable regime, in which either a threshold-like enhancement
or reduction of the local nuclear field by up to 3 Tesla can be generated by
varying the intensity of light. The excitation power threshold for such a
nuclear spin "switch" is found to depend on both external magnetic and
electric fields. The switch is shown to arise from the strong feedback of
the nuclear spin polarization on the dynamics of spin transfer from
electrons to the nuclei of the dot.\\
\end{abstract}

\maketitle

The hyperfine interaction in solids \cite{Overhauser} arises from the
coupling between the magnetic dipole moments of nuclear and electron spins.
It produces two dynamical effects: (i) inelastic relaxation of electron spin
via the "flip-flop" process (Fig.1a) and (ii) the Overhauser shift of the
electron energy \cite{nuclearbook}. Recently, the hyperfine interaction in
semiconductor quantum dots (QDs) has attracted close attention
\cite{Gammon,Bracker,Johnson,Petta,Koppens,Yokoi,Braun,Lai,Eble,Erlingsson,Khaetskii,Imamoglu} fuelled by proposals for QD
implementation in quantum information applications \cite{Loss}. The full
quantization of the electron states in QDs is beneficial for removing
decoherence mechanisms present in extended systems \cite{Kroutvar,Greilich}.
However, the electron localization results in a stronger (than in a bulk
material) overlap of its wave-function with a large number of nuclei ($N\sim
10^{4}$ in small self-assembled InGaAs/GaAs dots and up to $10^{5}\div
10^{6}$ in electrostatically-defined GaAs QDs), and the resulting hyperfine
interaction with nuclear spins has been found to dominate the decoherence
\cite{Johnson,Petta,Koppens,Erlingsson,Khaetskii,Imamoglu} and life-time \cite{Braun} 
of the electron spin at low temperatures.

In this Letter, we report the observation of a pronounced bistable behaviour
of nuclear spin polarisation, $S$, in optically pumped self-assembled
InGaAs/GaAs dots. In our experiments, spin-polarized electrons are
introduced one-by-one into an individual InGaAs dot at a rate $w_{x}$ (see
Fig.1b) by the circularly polarized optical excitation of electron-hole
pairs 120 meV above the lowest QD energy states. Due to hole spin-flip
during its energy relaxation, both bright and dark excitons can form in the
dot ground state. The former will quickly recombine radiatively with a rate
$w_{rec}\approx 10^{9}$ sec$^{-1}$, whereas the dark exciton can recombine
with simultaneous spin transfer to a nucleus via a spin "flip-flop" process
(as in Fig.1a) at the rate $w_{rec}Np_{hf}$ \cite{Erlingsson,GammonPRL}.
Here $N$ is the number of nuclei interacting with the electron and $p_{hf}$
is the probability of a "flip-flop" process, which from our perturbation
theory treatment is given by:
\begin{equation}
p_{hf}=|h_{hf}|^{2}/(E_{eZ}^{2}+\textstyle\frac{1}{4}\gamma ^{2}).
\label{Eq-Ws}
\end{equation}%
Here $\gamma $ is the exciton life-time broadening, $h_{hf}$ is the strength
of the hyperfine interaction of the electron with a single nucleus and
$E_{eZ}$ is the electron Zeeman splitting. $E_{eZ}$ is strongly dependent on
the effective nuclear magnetic field $B_{N}$ generated by the nuclei. This
provides a feedback mechanism between the spin transfer rate and the
degree of nuclear polarization ($B_{N}\propto S$) in the dot \cite{Tarucha}.
The feedback gives rise to bistability in the nuclear polarization and
threshold-like transitions between the spin states of $10^{4}$ nuclei
leading to abrupt changes of $B_{N}$ by up to 3T in few nanometre sized QDs.

\begin{figure}
\includegraphics[width=7cm]{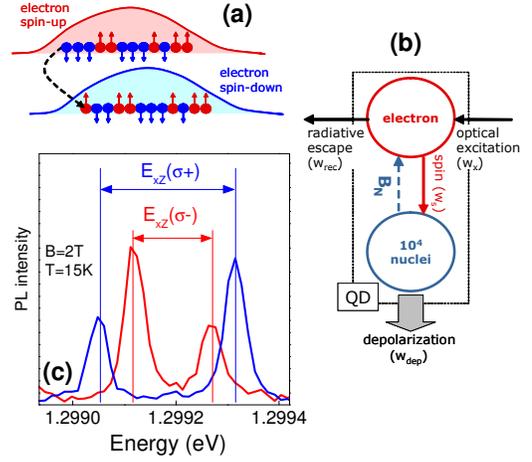}
\caption{(a) Schematic representation of the electron-nuclear spin "flip-flop"
process. (b) The steps involved in nuclear polarization of a quantum dot (see text
for detail). (c) $X^{0}$ photoluminescence
spectra recorded for an individual InGaAs QD in an external magnetic field
$B=2$T at $T=15$K. The spectrum excited with $\sigma ^{+}$ ($\sigma ^{-}$)
light resonant with the wetting layer is plotted in blue (red). The
horizontal arrows show the corresponding exciton Zeeman
splittings.}
\label{fig1}
\end{figure}

We observe such threshold-like transitions (referred to below as a nuclear
spin 'switch') in several different structures containing self-assembled
InGaAs/GaAs QD with $\sim $3x20x20 nm size. Below, we present results
obtained at a temperature of 15K for two GaAs/AlGaAs Schottky diodes, where
the dots are grown in the intrinsic region of the device. In these structures a
bias can be applied permitting control of the vertical electric field, $F$
\cite{structure}. For photoluminescence (PL) experiments, individual dots
are isolated using 800 nm apertures in a gold shadow mask on the sample
surface.

Fig.1c shows time-averaged (60s) PL spectra recorded for a neutral exciton
in a single QD in an external magnetic field of 2T. Circularly polarized
laser excitation in the low energy tail of the wetting layer (at 1.425eV) is
employed and unpolarized PL from the dot is detected using a double
spectrometer and a CCD. For each excitation polarization a spectrum
consisting of an exciton Zeeman doublet is measured with the high (low)
energy component dominating when $\sigma ^{+}$ ($\sigma ^{-}$) polarization
is used. A strong dependence of the exciton Zeeman splitting $(E_{xZ})$ on
the polarization of the excitation is observed in Fig.1c:
$E_{xZ}(\sigma^{+})=260\mu $eV and $E_{xZ}(\sigma ^{-})=150\mu $eV. Such a
dependence is a signature of dynamic nuclear polarization
\cite{Gammon,Bracker,Yokoi,Lai,Eble}, which gives rise to the
Overhauser field $B_{N}$ aligned parallel or anti-parallel
to $B$ for $\sigma ^{+}$ or $\sigma^{-}$ excitation, respectively.
\begin{figure}
\includegraphics[width=5.5cm]{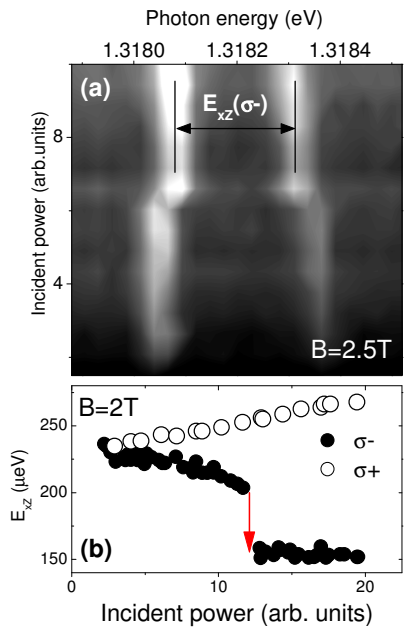}
\caption{(a) Grey-scale plot showing exciton PL spectra recorded for an
individual InGaAs dot. The spectra are recorded at $B$=2.5T using
unpolarized detection and $\sigma ^{-}$ excitation into the wetting layer.
The spectra are displaced along the vertical axis according to the
excitation power at which they are measured. (b) $E_{xZ}$ power dependences
measured at $B=2$T for $\sigma ^{+}$ and $\sigma ^{-}$ excitation
polarizations.}
\label{fig2}
\end{figure}

The dependence of exciton PL at $B=2.5$T on the power, $P\propto w_{x}$, of
$\sigma ^{-}$ excitation is shown in the grey-scale plot in Fig.2a. At low $P$
the Zeeman splitting $E_{xZ}=310\mu $eV. As the power is increased, a
threshold-like decrease of $E_{xZ}$ to $225\mu $eV is observed at $P=P_{up}$
indicating the sudden appearance of a large nuclear field. Fig.2b shows the
power dependence of $E_{xZ}$ measured at $B=2$T for both circular
polarizations of incident light. For $\sigma^{-}$ excitation, $E_{xZ}$
decreases below the threshold followed by a weak power dependence
above the threshold. The $\sigma^-$ behaviour contrasts the
weak monotonic increase of $E_{xZ}$ seen for $\sigma ^{+}$
excitation over the whole range of powers similar to that
reported in Ref.\cite{Brown}.

\begin{figure}
\includegraphics[width=6cm]{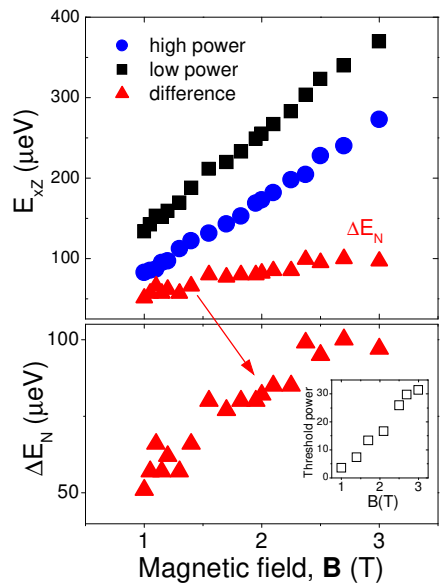}
\caption{Dependence of the QD exciton Zeeman splitting $E_{xZ}(\sigma ^{-})$ on
the external magnetic field. Squares and circles show high and low power
data, respectively, and triangles show their difference, $\Delta E_{N}$.
For all $B$ shown in the figure the nuclear switch threshold was observed with the
threshold power shown in the inset as a function of $B$.}
\label{fig3}
\end{figure}

The variation of the Zeeman splitting in Fig.2 reflects the change in the
nuclear field $B_{N}$:
$E_{xZ}(\sigma^{\pm })=|g_{e}+g_{h}|\mu_{B}B\pm|g_{e}|\mu _{B}B_{N}(\sigma ^{\pm })$
(where $g_{e}$ is the electron
$g$-factor \cite{holes}, $\mu _{B}$ is the Bohr magneton). $B_N$ in its turn
depends on the external field $B$. Triangles in Fig.3 show the difference
between $E_{xZ}(\sigma ^{-})$ at low and high powers (squares and circles in
Fig.3a, respectively), $\Delta E_{N}=|g_{e}|\mu _{B}B_{N}(\sigma ^{-})$, as
a function of $B$. $\Delta E_{N}$ increases linearly with $B$ at low fields
and then saturates at $B\approx 2.5\div 3$T. The inset in Fig.3 shows that
the threshold-power for the switch also increases nearly linearly with $B$.
No switch could be observed at $B>3$T in the range of powers employed in our
studies.

For $B<3$T, when the excitation power was gradually reduced from powers
above the switch, $E_{xZ}$ was found to vary weakly with power until another
threshold was reached, where the magnitude of the exciton Zeeman splitting
abruptly increased (at $P=P_{down}$), as shown in Fig.4. This increase of
$E_{xZ}$ corresponds to depolarization of the nuclei and hence reduction of
$B_{N}$. The observed hysteresis of nuclear polarization shows that two
significantly different and stable nuclear spin configurations can exist for
the same external parameters of magnetic field and excitation power. We find
that high nuclear polarization persists at low excitation powers for more
than 15 min, this time most likely being determined by the stability of the
experimental set-up.

We also show in Figs.4a,b that the size of the hysteresis loop depends on
the external magnetic or electric fields (the electric field is given by
$F=(V_{rev}+0.7V)/d$, where $V_{rev}$ is the applied reverse bias and $d=230nm
$ is the width of the undoped region of the device). The inset in Fig.4b
shows the $P_{up}$ reverse bias-dependence. In general, both $P_{up}$ and
$P_{down}$ increase with $B$ and reverse bias, but also the difference
between the two thresholds increases, leading to a broader range of incident
powers in which the bistability occurs. The threshold bias dependence arises
from the influence of the electric field on the charge state of the dot
\cite{structure}, and will be discussed elsewhere.

In order to explain the nuclear switching and bistability, we employ a model
based on spin-flip assisted e-h recombination \cite{Erlingsson,GammonPRL}.
We assume that the electron spin is defined by the sign of the circularly
polarized excitation ($\sigma^{\pm}$), whereas the hole spin is partially
randomized during the energy relaxation. Thus, dark and bright excitons
can be formed in the dot ground state, with the rates $\alpha w_{x}$ and
$(1-\alpha )w_{x}$, respectively. A bright exciton recombines with the rate
$w_{rec}$ without spin transfer to the nuclei. In contrast, a dark exciton
can recombine with the electron simultaneously flipping its spin due to the
hyperfine interaction: the electron virtually occupies an optically active
state with the opposite spin and the same energy \cite{Erlingsson,GammonPRL}
transferring spin to nuclei and, then, recombines with the hole with the
rate $w_{rec}Np_{hf}$, where $p_{hf}$ (given by Eq.(\ref{Eq-Ws}) depends on
the electron Zeeman splitting,
$E_{eZ}=|g_{e}|\mu _{B}[B\pm B_{N}(\sigma^{\pm })]$.
For the case of $\sigma^{-}$ excitation, polarization of the
nuclei leads to a decrease of $E_{eZ}$, and thus a positive feedback and
speeding up of the spin transfer process: the more spin is transferred to
the nuclear system the faster becomes the spin transfer rate. By contrast
for $\sigma^{+}$ excitation, spin transfer leads to an increase of $E_{eZ}$,
leading to the saturation of $S$ (and $B_{N}$) at high power.

The spin transfer to nuclei at a rate $w_{s}\propto \alpha w_{x}Np_{hf}$
competes with nuclear depolarization, $\dot{S}=-w_{dep}S$ (see Fig.1b)
due to spin diffusion away from the dot into the surrounding GaAs
\cite{Paget,Wdep}, at a rate $w_{dep}\sim 1\div 10$s$^{-1}$. At high power of
$\sigma ^{-}$ excitation $w_{s}$ may exceed $w_{dep}$ and then a stimulated
nuclear polarization will take place due to the positive feedback mechanism
described above leading to an abrupt increase of the nuclear spin
(at $P=P_{up}$). To achieve the condition $w_s=w_{dep}$ a higher
$w_x$ (power) will be required at higher $B$ in agreement with
observation in the inset of Fig.3. The stimulation at
$P \approx P_{up}$ stops when either (i) $|E_{eZ}|$ starts
increasing again since $B_{N}>B$, causing reduction of $w_{s}$ or (ii) the
maximum achievable $B_{N}=B_{N}^{max}$ in the given dot is reached. This
explains the dependence in Fig.3, where $\Delta E_{N}$, and hence the
nuclear field, increases at low $B$ and saturates at high fields, from which
we estimate $B_{N}^{max}\approx $2.5-3T \cite{gfactor}.

When the power is reduced from beyond the threshold $P_{up}$, and the
condition $w_{s}<w_{dep}$ is reached at sufficiently low $w_{x}$, a strong
negative feedback is expected: further nuclear depolarization will lead to
even lower $w_{s}$ due to the increase in the electron Zeeman energy
$E_{eZ}$. Thus, an abrupt nuclear depolarization will take place (at the threshold
$P_{down}$). This explains the observed hysteresis behavior in Fig.4, and
also accounts for the existence of a bistable state in the nuclear
polarization at intermediate powers, $P_{down}<P<P_{up}$.

\begin{figure}
\includegraphics[width=6cm]{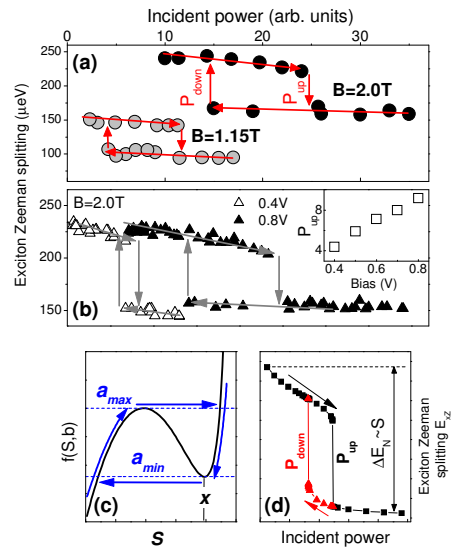}
\caption{(a) Power dependence of $E_{xZ}(\sigma ^{-})$ measured at $B=2$T and
1.15T. The arrows show the direction in which the hysteresis loop is
measured with two thresholds $P_{up}$ and $P_{down}$. (b)
$E_{xZ}(\sigma^{-})$ power dependence measured at $B=2.0$T. The two hysteresis loops are
measured at 0.4 and 0.8V applied bias. The inset shows the $P_{up}$
dependence on the reverse bias applied to the diode. (c) The full line shows
the function $f(S,b)$ from Eq.\ref{Sn}. Arrows show how the
hysteresis loop is formed when the parameter $a$ ($\propto w_{x}$)
is varied for a fixed $b$. (d) Hysteresis loop of the exciton Zeeman splitting as a function of
incident power calculated using Eq.\ref{Sn} for $x=0.7$
and $\theta=0.1$.}
\label{fig4}
\end{figure}

To model this bistability, we solve the rate equations for the nuclear spin
polarization $S$, and populations of bright and dark excitons, $n_{b}$ and
$n_{d}$, ($1-n_{b}-n_{d}$ is the probability that the dot is empty):
\begin{eqnarray*}
\dot{S} &=&n_{d}w_{rec}p_{hf}(1-S)-w_{dep}S, \\
\dot{n}_{b} &=&\left( 1-\alpha \right) w_{x}[1-n_{b}-n_{d}]-w_{rec}n_{b},\\
\dot{n}_{d} &=&\alpha w_{x}[1-n_{b}-n_{d}]-\frac{1}{2}(1-S)Nw_{rec}p_{hf}n_{d}.
\end{eqnarray*}%
In the limit $\gamma \ll |g_{e}|\mu _{B}B_{N}^{max}$ we obtain the
following equation for a steady state polarisation
induced by the $\sigma ^{\pm }$ excitation:
\begin{equation}
f(S,b)\equiv S\left[ 1+b\frac{(x\pm S)^{2}}{1-S}\right] =a,\;\;x=\frac{B}{%
B_{N}^{max}},  \label{Sn}
\end{equation}%
where for $w_{x}\ll w_{rec}$ (low occupancy of the dot)
\begin{equation}
a=2\alpha w_{x}/Nw_{dep}\quad ,\quad b=2\alpha Nw_{x}/w_{rec}.  \label{ab}
\end{equation}

In Eq.(\ref{ab}), both $a$ and $b$ are proportional to the excitation power.
For low excitation powers such that $b\ll 1$, for both $\sigma ^{+}$ and
$\sigma ^{-}$ excitation, Eq.(\ref{Sn}) has a single solution for the degree
of nuclear polarization, namely $S\approx a$. In the $\sigma ^{+}$
excitation case, $f(S,b)$ is a monotonic function and for all $a$ and $b$ a
single solution to Eq.(\ref{ab}) is obtained. On the other hand, for $\sigma
^{-}$ excitation, for higher powers such that $b\gtrsim 1$, $f(S,b)$
acquires an N-shape, as illustrated in Fig.4c. As shown in the diagram, an
abrupt transition to $S>x$ ($S\approx a$) will be obtained when $a_{max}$
($a_{min}$) is reached at the local maximum (minimum) of $f(S,b)$. The
transitions at $a_{max}$ and $a_{min}$ correspond to the $P_{up}$ and
$P_{down}$ thresholds in Fig.4, respectively, whereas for
$a_{min}<a<a_{max}$, the polarization degree $S$ enters a regime of
bistability in which the cubic Eq.(\ref{Sn}) has three solutions, two
of which are stable with an unstable one in between.

We find that the occurrence of the switch to $S>x$ depends on the
dimensionless ratio $\theta =a/b=w_{rec}/N^{2}w_{dep}$, since at small
$\theta $, $a$ will grow more slowly with $w_{x}$ than the magnitude of $f(S,b)$
at the local maximum. $\theta $ is determined by the dot parameters only,
and can be estimated for the dots studied in our experiment: we obtain
$\theta _{exp}\sim 1\div 10$ from $w_{rec}\sim 10^{9}$ sec$^{-1}$,
$w_{dep}\sim 1\div 10$ sec$^{-1}$ and $N\sim 10^{4}$. Using Eq.(\ref{Sn}) we
find that for $x\leq 0.8$ the spin switch is possible for any
$\theta_{exp}>\theta _{c}$, where
$\theta_{c}=\frac{1}{16}(3-\sqrt{9-8x})(4x-3+\sqrt{9-8x})^{2}/(1+\sqrt{9-8x})\leq 0.1$,
consistent with our observations.
A hysteresis loop calculated using Eq.(\ref{Sn}) for $x=0.7$ (with
$\theta_{c}\approx 0.07$) and $\theta =0.1$ (close to critical
$\theta _{c}$), is
shown in Fig.4d.

To summarize, we have observed a strong optically induced bistability of the
nuclear spin polarization in self-assembled InGaAs QDs. We show that nuclear
magnetic fields up to 3T can be switched on and off in individual dots by
varying one of three external controlling parameters: electric and magnetic
fields and intensity of circularly polarized excitation. We have found that
the nuclear spin switch effect is a general phenomenon and has been observed
in several different InGaAs/GaAs quantum dot samples at temperatures
$T=15-30$K and in the range of external magnetic fields $B=1\div 3$T. The effect
arises due to the strong feedback of the nuclear spin polarization on the
dynamics of the electron-nuclear spin interaction accompanying the radiative
recombination process, which is enhanced when the Overhauser and external
magnetic fields cancel each other.

We thank R. Oulton, A. Imamoglu, C. Marcus, D. Loss, and S. Tarucha for
discussions. This work has been supported by the Sheffield EPSRC Programme
grant GR/S76076, the Lancaster-EPSRC Portfolio Partnership EP/C511743, the
EPSRC IRC for Quantum Information Processing, ESF-EPSRC network EP/D062918,
and by the Royal Society. AIT acknowledges support from the EPSRC Advanced
Research Fellowship EP/C54563X/1 and research grant EP/C545648/1.


\begin{references}
\bibitem{holes} The hyperfine interaction is negligible for holes due to the p-like nature of the Bloch functions.
\bibitem{Wdep} For a dot with the radius $r\sim 5$nm we estimated
$w_{dep}\sim D_{N}/r^{2}\sim \mu _{n}^{2}/(\hbar ar^{2})\sim 1\div 10$s$^{-1}$,
where $D_{N}\sim \mu _{n}^{2}/\hbar a$ is the coefficient of
polarization diffusion due to the dipole-dipole interaction
($\sim \mu_{n}^{2}/a^{3}$) between magnetic moments
$\mu _{n}$ of neighbouring nuclei and $a=0.56$nm - lattice constant.
\bibitem{gfactor} Note that a close value of $B_{N}$ was deduced from the
dependence of $\Delta E_{N}(B)$ using $|g_{e}|\approx 0.5$.
\end{references}
\end{document}